\documentclass[%
 reprint,
 amsmath,amssymb,
 aps,
]{revtex4-2}

\usepackage{inputenc} 
\usepackage{fontenc}
\usepackage[french,english]{babel}
\usepackage{geometry} 
\usepackage{hyperref} 
\usepackage{bookmark}
\usepackage[normalem]{ulem} 
\usepackage{soul} 
\usepackage{color} 
\usepackage[table]{xcolor} 
\usepackage{float} 
\usepackage{hyphenat} 
\usepackage{setspace} 
\usepackage{url} 

\usepackage{verbatim} 
\usepackage{dcolumn}
\usepackage{graphicx} 
\usepackage{subfigure}

\usepackage{amsmath} 
\usepackage{amssymb} 
\usepackage{bm}
\usepackage{siunitx}
\usepackage{lineno} 
\makeatletter
\newcommand{\myfnsymbol}[1]{%
  \expandafter\@myfnsymbol\csname c@#1\endcsname
}
\newcommand{\@myfnsymbol}[1]{%
  \ifcase #1
  \or \TextOrMath{\textasteriskcentered}{*}
  \or $\dag$
  \or $\ddag$
  \or $\P$
  \fi
}
\newcommand{\equalcontributor}{\@myfnsymbol{1}}
\newcommand{\affiliationA}{\@myfnsymbol{2}}
\newcommand{\affiliationB}{\@myfnsymbol{3}}
\newcommand{\affiliationC}{\@myfnsymbol{4}}
\makeatother

\begin{document}

\preprint{AIP/123-QED}

\title{Homodyne detection for pulse-by-pulse squeezing measurement}

\author{Tiphaine Kouadou}
  \altaffiliation[Currently at ]{Illinois Quantum Information Science and Technology Center, Department of Physics, The Grainger College of Engineering, University of Illinois Urbana-Champaign, Urbana, IL 61801, USA}
\author{Elie Gozlan}
\author{Lo\"ic Garcia}
\author{David Polizzi}
\author{David Fainsin}
\author{Iris Paparelle}
\author{R. L. Rinc\'on Celis}
\author{Bastien Oriot}
\author{Anthony Abi Aad}
\author{Peter Namdar}
\author{Gana\"el Roland}
\author{Nicolas Treps}
\author{B\'ereng\`ere Argence}
\author{Valentina Parigi}
\email{valentina.parigi@lkb.upmc.fr}

\affiliation{Laboratoire Kastler Brossel, Sorbonne Universit\'e, CNRS, ENS-PSL Research University, Coll\`ege de France\\4 place Jussieu, F-75252, Paris, France}


\begin{abstract}
\noindent
Homodyne detection is a phase-sensitive measurement technique, essential for the characterization of continuous-variable (CV)-encoded quantum states of light. It is a key component to the implementation of CV quantum-information protocols and benefits from operating, by design, at room temperature. However, performing high-speed quantum information processing remains a major challenge, as conventional homodyne detectors often fail to sustain pulsed operation at high repetition rates due to electronic limitations. We present wideband homodyne detectors operating at near-infrared (NIR) and telecom wavelengths, with optimized performance at repetition rates up to 150 MHz. We demonstrate their performance by resolving the pulse-by-pulse structure of squeezed states of light at telecom wavelengths while preserving their spectral multimode properties.
\end{abstract}


\renewcommand{\thefootnote}{\myfnsymbol{footnote}}
\maketitle


\footnotetext[1]{tiphaine@illinois.edu. \\ Current address: Department of Physics, University of Illinois Urbana-Champaign, Urbana, IL 61801, USA}

\footnotetext[2]{valentina.parigi@lkb.upmc.fr}%

Coherent detection lies at the core of photonics technologies.
In the quantum regime, homodyne detection is a key element for secure communication \cite{Hajomer24,Pan22,Pietri24}, sensing \cite{Domeneguetti23,Stefszky23}, and computation \cite{Asavanant24} at room temperature. Specifically, if information is encoded in the Continuous Variables (CV) of the field, and a reliable supply of quantum states with non-Gaussian statistics is provided, computational operations can be efficiently performed via homodyne setups \cite{Bourassa2021}.
Homodyne detectors, known for achieving high quantum efficiencies at room temperature, have been further developed to meet specific requirements, e.g. bandwidth, speed or dynamic range, according to the specific characteristics of the optical system supporting the quantum resource \cite{Kumar2012,Ng2024}.

Large-bandwidth detectors are crucial to process large-bandwidth information. Alternative options to finite-bandwidth homodyne detection have been developed, such as detection by parametric amplification \cite{Inoue2023, Tian22,Takanashi20, Shaked2018} or unbalanced detection \cite{Wilken24}. However, the first method requires the gain of the parametric-amplification process to be larger than the gain with which the signal of interest -- non classical -- is generated, while the second requires the complete suppression of the noise of the laser used as a local oscillator.   
 
In this work, we present homodyne detectors capable of operating on NIR and telecom pulsed sources at repetition rates of up to 150 MHz. To showcase their potential, we resolve the pulse-by-pulse structure of squeezed telecom light, while preserving the, recently demonstrated, spectrally multimode architecture of such sources \cite{RomanRodriguez2024}.

Operating homodyne detectors in the pulsed regime presents significant challenges. The high-frequency part of the detected signals strains the detector electronics, thereby limiting the experimentally achievable bandwidth. Consequently, most commercially available detectors cannot sustain high-repetition-rate pulsed operations, even when they use large-bandwidth electronic components. Here, we report on the circuit design and electronic components selected to enhance the detection bandwidth while maintaining the shot-noise clearance (SNC) required to operate in the quantum regime.

\section{Quadrature measurement with homodyne detection}
\subsection{Ideal detection}
\begin{figure}[h!]
\centering
\includegraphics[width=\linewidth]{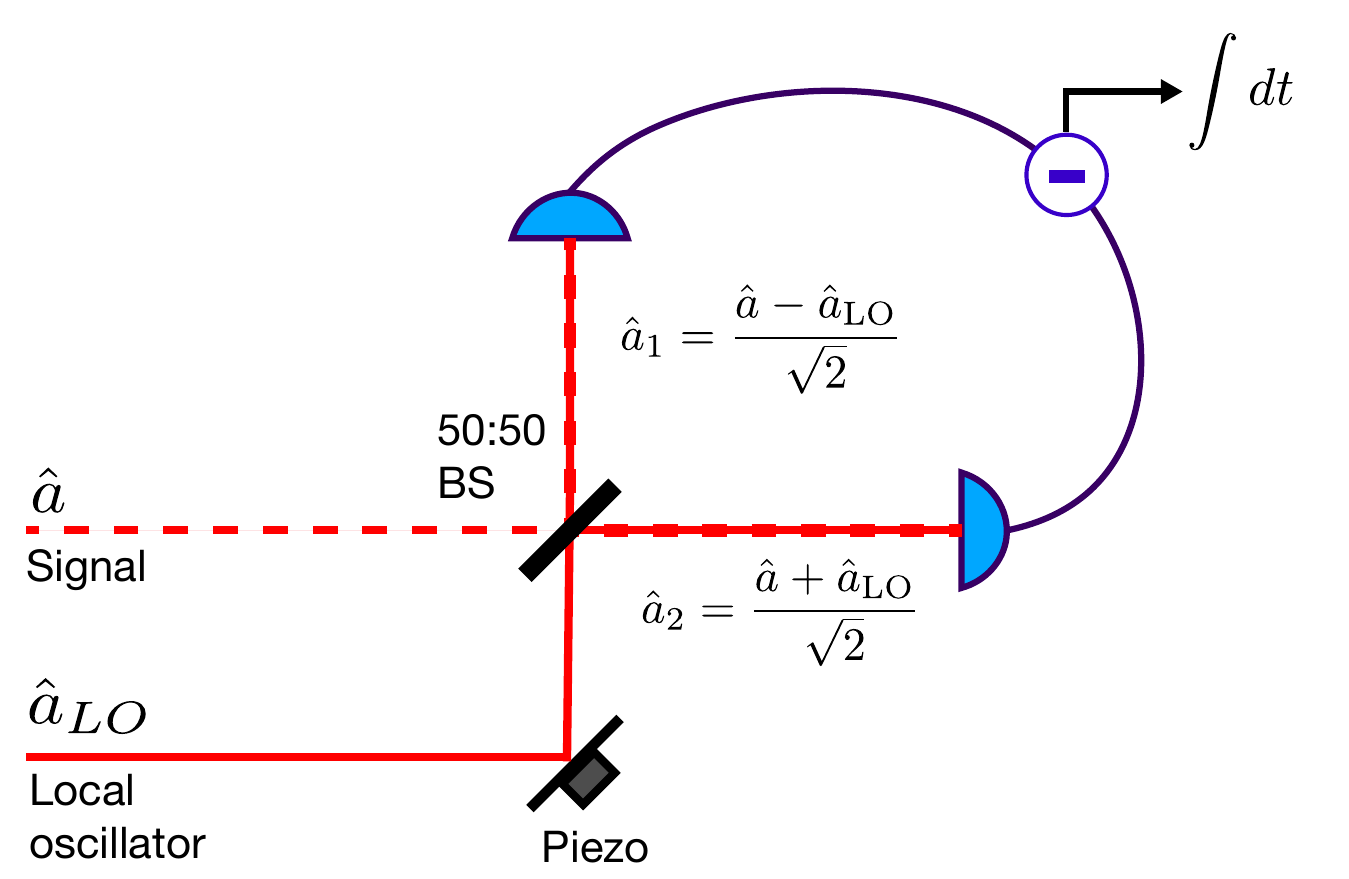}
\caption{\small{Schematics of a homodyne setup. BS: beam splitter; Piezo: piezo-electric actuator.}}
\label{fig:simplehomo}
\end{figure}

Homodyne detection (HD) is a measurement technique based on the interference between a beam of interest, the signal, and a reference beam called Local Oscillator (LO), usually derived from the same laser source. In a two-port homodyne \cite{Schumaker1984,Yuen1983}, light is mixed at a lossless and balanced (50/50) beam splitter (BS), and the outgoing beams are then sent to two photodiodes (PD) (see Fig. \ref{fig:simplehomo}). 

 
The light detected by each PD is converted into photocurrents $i_1$ and $i_2$, that are subsequently subtracted to generate the homodyne signal: 
$\hat{N}_{21} = \hat{N}_2 \, - \, \hat{N}_1 = \hat{a}^{\dag}_{LO}\hat{a} \, + \, \hat{a}^{\dag}\hat{a}_{LO}$, where $\hat{a}$ and $\hat{a}_{LO}$ are the annihilation operators for the signal of interest and the LO field respectively.
In the case of a strong LO, i.e., $\langle \hat{a}_{LO}^{\dag}\hat{a}_{LO} \rangle \gg \langle  \hat{a}^{\dag}\hat{a} \rangle$, and negligible LO fluctuations compared to the mean field, the bosonic operators of the local oscillator can be replaced with their mean value $\alpha_{LO} = \lvert  \alpha_{LO} \rvert e^{- \imath \theta }$, i.e., the field amplitude. Thus, $\hat{N}_{21}$ can be approximated by

\begin{equation}
\hat{N}_{21} = \lvert  \alpha_{LO} \rvert  \, \hat{q}_{\theta},
\label{eq:hd_signal_strongLO}
\end{equation}

where the amplitude of the signal depends on that of the LO, and $\hat{q}_\theta = \hat{a} e^{-i\theta} \, + \, \hat{a}^{\dag} e^{i\theta} = \hat{q} \, \cos(\theta) + \hat{p} \sin(\theta)$ is the general field quadrature. The angle $\theta$ describes the phase shift of the quadratures $\hat{q}$ and $\hat{p}$ in the phase space, such that $\hat{q}_{\theta = 0} = \hat{q}$ and, $\hat{q}_{\theta = \frac{\pi}{2}} = \hat{p}$.
By scanning the phase of the LO, i.e., changing $\theta$, we can access the quadratures in the whole phase space, as well as derived quantities, i.e., the mean quadrature $\langle \hat{N}_{21} \rangle = \alpha_{LO} \, \langle \hat{q}_{\theta} \rangle$ or the variance:

\begin{equation}
\Delta^2 \hat{N}_{21} = \lvert \alpha_{LO} \rvert  ^2 \; \Delta^2 \hat{q_{\theta}}. \label{eq:subtraction_variance}
\end{equation} 

This is the quantity that we compare to the shot-noise fluctuation: when the quadratures fluctuations are less than that of the shot noise, the state quadrature is \textit{squeezed}. When the fluctuations are higher, the state is \textit{anti-squeezed}. 


\subsection{Real Photodetection}
In practice, the recovered homodyne signal $\mathcal{N}_{21}^{\tau}$ results from the subtraction of the photocurrents $i_j$ generated by each detector and integrated over the response time of the detector $\tau$\cite{Cooper2013,Raymer1995}:
\begin{equation}
\mathcal{N}_{21}^{\tau} = \int_0^{\tau} dt \, \left( \hat{i}_2  - \hat{i}_1 \right).
\label{eq:integrated_signal}
\end{equation}

Therefore, to measure the signal of a single pulse, the integration time must be shorter than the period between consecutive pulses, i.e., the inverse of the laser repetition rate. \\

To model a realistic homodyne detection, we describe losses as undesired vacuum contributions combined to the noiseless homodyne signal with a BS, which splitting ratio is now $\sqrt{\eta} \, /\sqrt{1 - \eta}$ \cite{Leonhardt1997}. In this case, the homodyne signal becomes:
\begin{equation}\label{eq:vacHD}
\hat{N'}_{21}= \lvert \alpha_{LO} \rvert(\sqrt{\eta} \, \hat{q}_{\theta} + \sqrt{1-\eta} \; \hat{q}_{0,\theta} ) 
\end{equation}

with $\eta \in [0,1]$. A more detailed derivation of the homodyne signal in the case of a non-ideal photodetection can be found in \cite{Kouadou2021}. The overall detection efficiency $\eta$ depends on the photodiode efficiency, the electronic noise of the detector and the quality of the interference between the beams before the photodiode such that:

\begin{equation}
\eta = \eta_{mod} \times \eta_{_{PD}}^2 \times \eta_{elec},
\label{eq:detector_efficiency}
\end{equation}

where $\eta_{mod}$ represents interferometric losses due to imperfect mode-matching between the signal and the local oscillator (LO) before the detector. $\eta_{_{PD}}$ is the photon-to-electron conversion efficiency of the photodiodes (two), and depends on the photosensitivity $S_{PD}$ (in A/W) of the PD material. Here, 
\begin{equation}
\eta_{_{PD}} (\lambda) = S_{PD} \, \frac{\lambda \, e}{h\,c},
\label{eq:QE_responsivity}
\end{equation}
where $e$ is the elementary charge, $h$ is the Planck constant, $c$ is the speed of light in vacuum, and $\lambda$ is the wavelength of the detected beam.

Finally, $\eta_{elec}$ is the electronic noise (EN) of the detector, i.e., the noise generated by the electrical components of the circuit and perturbations from the detector environment. It is defined as \cite{Kumar2012,Appel2007}


\begin{equation}
\eta_{elec} = 1 - \frac{\sigma_{en}^2}{\sigma_o^2},
\label{eq:elec_efficiency}
\end{equation}
where $\sigma_{en}^2$ is the variance of the measured EN and $\sigma_o^2$ is the variance of the output signal. In the next section, we will describe the major contributions to the electronic noise in our detector, and describe important parameters for the design and characterization of a HD, in particular, and how they can be optimized to increase the detector efficiency. 

\subsection{Performance Metrics of a Homodyne Detector}
The HD signal, being the difference between two similar photocurrent values, is usually a very small quantity, requiring electronic amplification to process the quantum signal. This also means that such devices saturate very easily when the incoming signal increases.
In particular, in a pulsed HD system, the incoming signal displays a large component at the laser repetition rate. Therefore, the most critical stage of detection is the subtraction of the photocurrent, especially at the laser repetition rate frequency. This requires precise alignment of the 50/50 BS and careful balancing of the two PD responses in the electronic circuit.


\noindent\paragraph*{\textbf{Common-mode rejection ratio}} To assess the performance of photocurrent subtraction, we compare the electronic signal produced by the circuit when only one photodiode is illuminated (common mode) and when  both PDs are illuminated. The quantity measured this way is called common-mode rejection ratio (CMRR) and is defined by:
\begin{align}
\text{CMRR} &= \frac{\sigma_{CM}^2}{\sigma_{d}^2} \label{eq:cmrr}
\end{align}

where $\sigma_{CM}$ denotes the common mode signal, i.e, the signal from one PD and $\sigma_d$ is the difference signal. To assess the performance of the HD, the power spectral density (PSD) of the common mode (CM) can be measured over the operating bandwidth of the detector, and the CMRR retrieved with eq. \ref{eq:cmrr}. With pulsed light, the PSD features a peak at the laser repetition rate $f_{rep}$ (see Fig. \ref{fig:muteq_cmrr_144}). In this case, we measure the CMRR at this frequency, i.e., where the subtraction performance is the most sensitive. \\



\paragraph*{\textbf{Shot-noise clearance}} Once the detector is aligned and calibrated, we measure its SNC (eq. \ref{eq:db_snc}), i.e, we compare the shot-noise level (signal of interest), to that of the dark noise (DN). The result in decibels (dB) is defined as
\begin{equation}
SNC_{dB} = 10 \times \text{Log}_{10} \left( \frac{\sigma_o^2}{\sigma_{en}^2} \right) \label{eq:db_snc}
\end{equation}

where $\sigma_o^2 = \sigma_{SN}^2 + \sigma_{en}^2$ denotes the output signal as the sum of the shot noise of the local oscillator $\sigma_{SN}^2$ and the electronic noise $\sigma_{en}^2$.
Thus, the amount of squeezing that can be measured depends on the SNC of the  homodyne detector.

\paragraph*{\textbf{Electronic noise floor}}
The EN of the detector is the accumulation of the noise generated by electrical components and perturbations in the detector environment. In our circuits, the main sources of electronic noise are the noise of the operational amplifiers and the thermal noise. \\ 
In the next section, we will describe the structure of the HD and discuss how the circuit design and the choice of its electronic components are crucial to the performance of the detector.


\begin{figure}
\centering
\includegraphics[width=8.5cm]{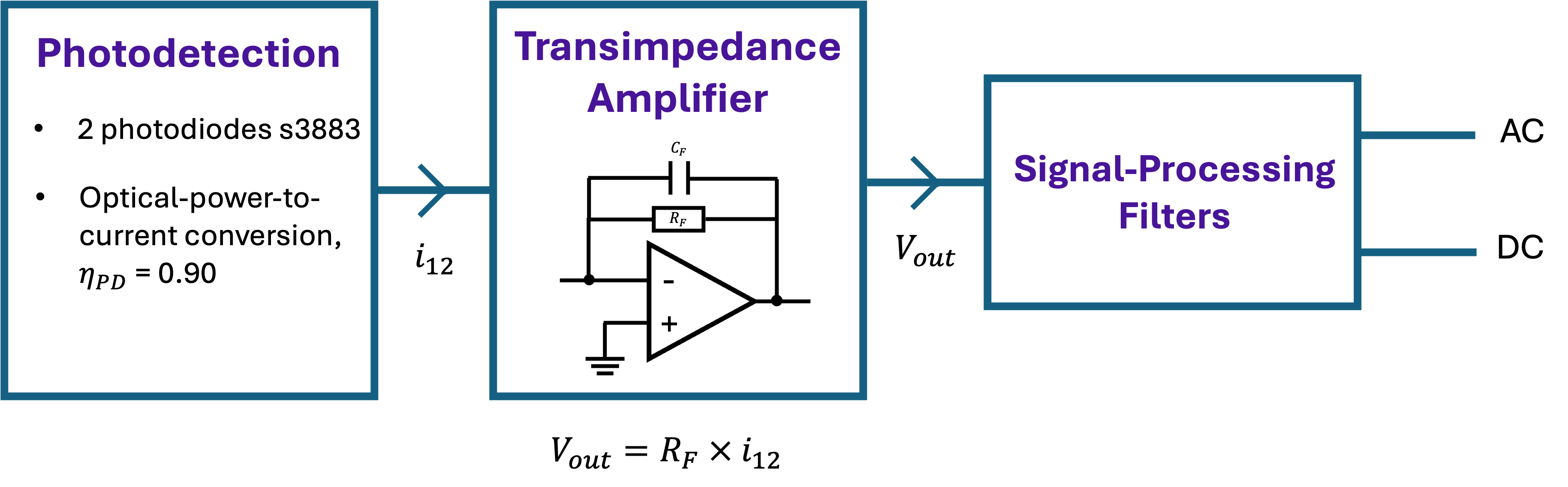}
\caption{Simplified schematics of a homodyne detector. $i_{12}$: subtracted photocurrent; $R_F$ and $C_F$ or the feedback resistance and capacitance respectively; $V_{out}$ is the voltage at the output of the TIA.}
\label{fig:basicHDscheme}
\end{figure}

\section{Structure of a homodyne detector}
\subsection{Photodetection}
Fig. \ref{fig:basicHDscheme} shows a general description of a HD, including a signal detection, amplification and demultiplexing (optional). We use Silicon (Si) photodiodes (PD) for the detection of NIR light and Indium Gallium Arsenide (InGaAs) PD for telecom detection. 

\begin{figure}
\centering
\includegraphics[width=0.35\linewidth]{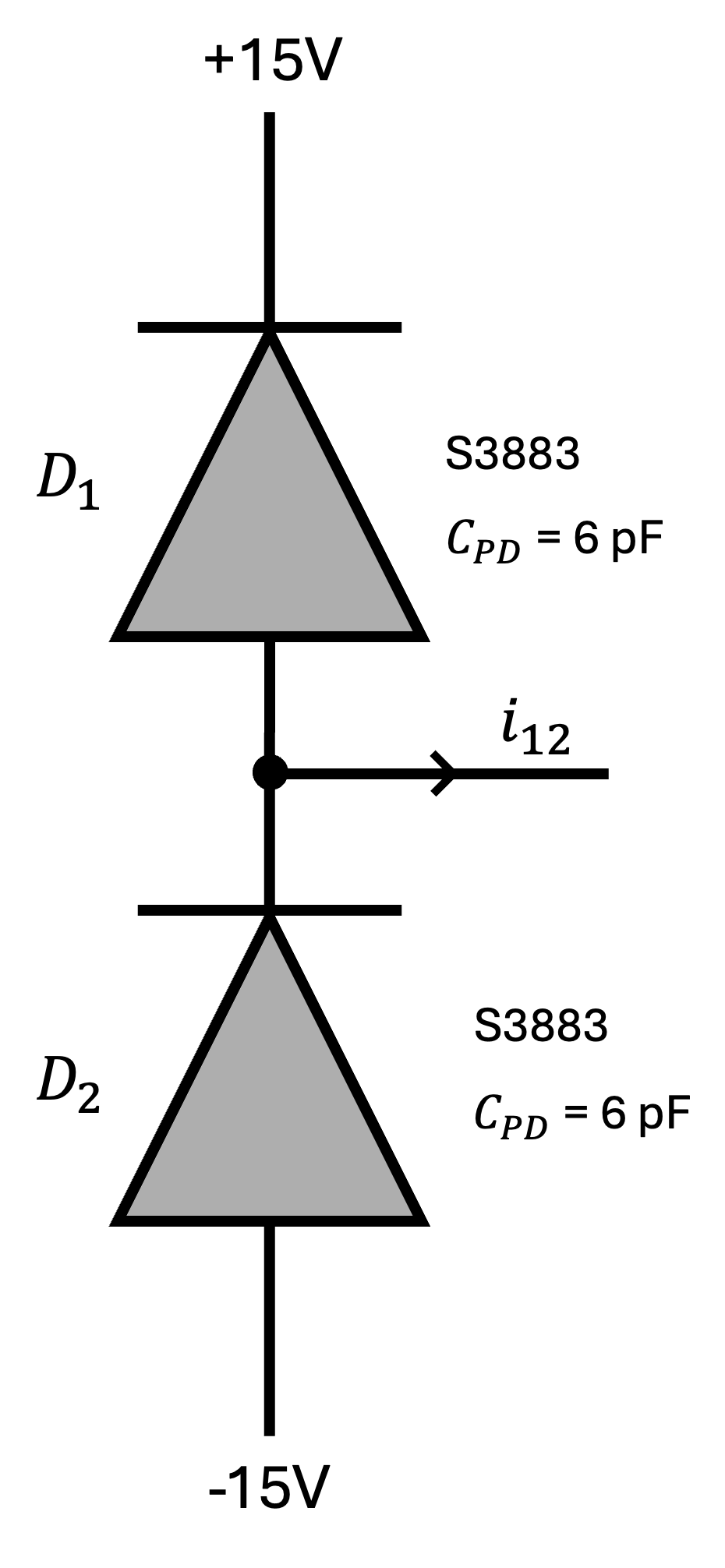} 
\caption{\label{fig:photodetection} Schematics of the photodedection and current subtraction. s3883: Hamamatsu Si photodiodes for the detection of NIR light.}
\end{figure}

In order to produce a homodyne electronic signal (eq. \ref{eq:subtraction_variance}), two PDs are mounted in reverse (see Fig. \ref{fig:photodetection}), to allow the subtraction of the photocurrents, $i_1$ and $i_2$, according to Kirchhoff's current law \cite{horowitz1989}. 

The PDs are connected to a power supply to apply a voltage bias. The voltage supply has an impact on the PD response; ideally, the electronic response over the specified bandwidth of the PD is flat. This aspect is critical to maintain a constant photosensitivity in the detection of photons over the operating range of the PD. Furthermore, the applied voltage is related to the effective intrinsic capacitance of PD $C_F$; for this reason it is critical for the capacitance of the PDs to remain identical throughout the  homodyne measurement. To ensure this is the case, we connect the PDs to two separate power supplies, which allows us to modify the bias voltage, and match the response of both PD. The optimal bias voltage corresponds to the best measurement of the CMRR. 

\subsection{Transimpedance Amplifier}
\begin{figure}
\centering
\includegraphics[width=5cm]{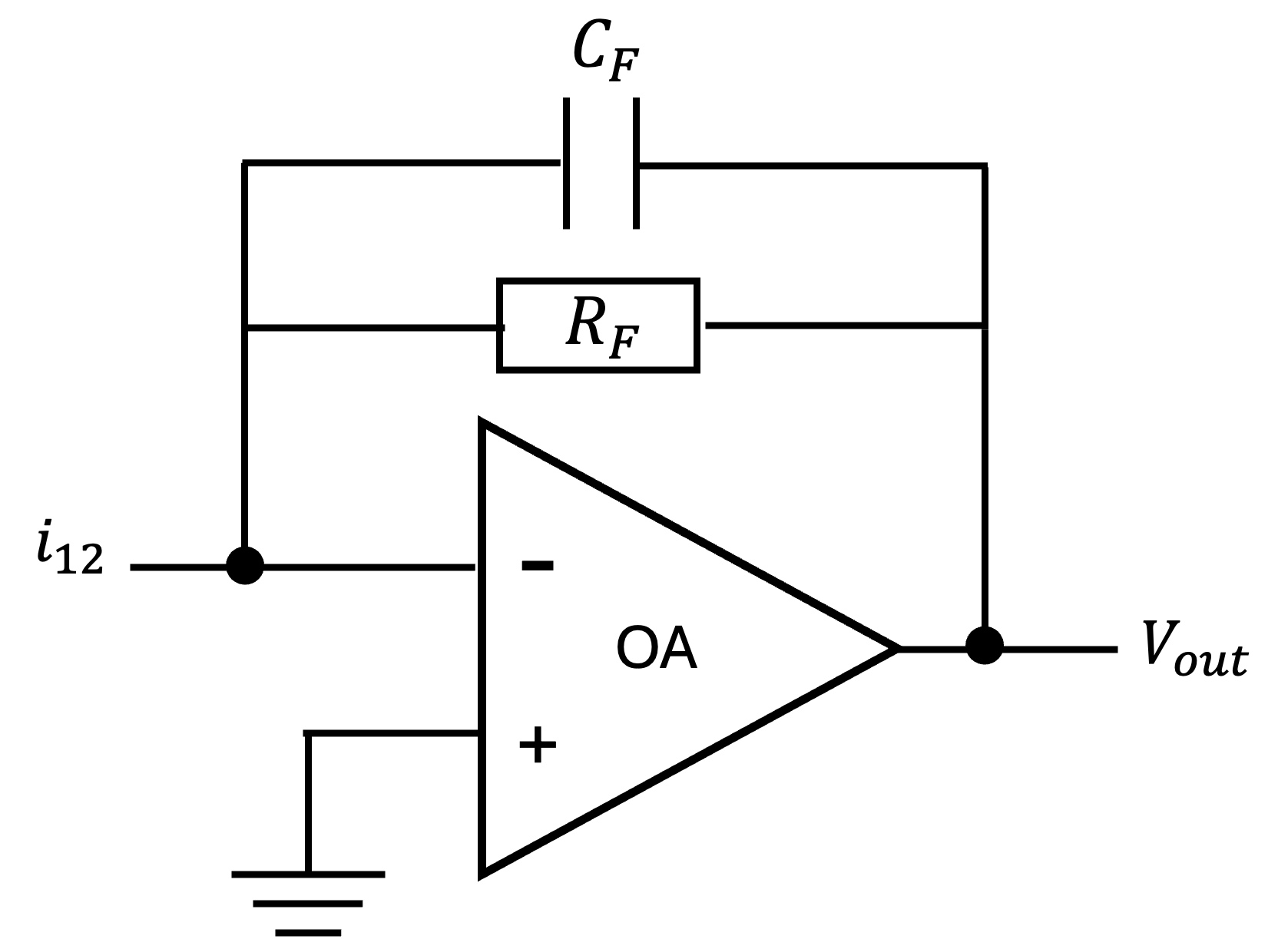}
\caption{\label{fig:TIAscheme}Simplified schematic of a transimpedance amplifier (TIA).}
\end{figure}

The produced subtracted photocurrent $i_{21}$ is sent to a Transimpedance Amplifier (TIA). It is a current-to-voltage converter with an adjustable gain (Fig. \ref{fig:TIAscheme}). It is composed of an operational amplifier (OA) connected to a feedback resistor $R_F$ and capacitor $C_F$. An OA is characterized by its gain-bandwidth product $GBP$, which indicates its performance. The gain of the current-to-voltage conversion is set by the feedback resistor $R_F$, the transimpedance gain, such that:

\begin{equation}
V_{_{TIA}} = R_F \times i_{_{21}}
\label{eq:V_TIA}
\end{equation}

where $V_{_{TIA}}$ is the voltage at the TIA output. The frequency response of the TIA (Bode plot \cite{horowitz1989}) depends not only on the OA and the feedback resistor, but also on the characteristics of the PDs. Thus, it is possible to estimate the signal cutoff frequency of the TIA with the following equation:
\begin{equation}
f_{_{TIA}} = \sqrt{\frac{GBP}{2\pi \, R_F \, (C_F \, + \, 2C_{PD} \, + \, C_{A1})}}
\label{eq:cutoff_tia}
\end{equation}

Here $C_{PD}$ is the intrinsic capacitance of the photodiodes and $C_{A1}$ the intrinsic capacitance of the operational amplifier, both indicated in the datasheet of the components. The feedback capacitor $C_F$, wired in parallel to $R_F$ (Fig. \ref{fig:TIAscheme}), allows to adjust the frequency response of the circuit, i.e., its stability. It is defined by \cite{Masalov2017}:

\begin{equation}
C_F = \sqrt{\frac{2 \, C_{PD} \, + \, C_{A1}}{\pi \times GBP \times R_F}}.
\label{eq:feedback_capacity_TIA}
\end{equation}

The signature of a stable homodyne circuit is a flat Bode plot over the detector's bandwidth followed by a gain decrease beyond the cutoff frequency (at -3 dB). If $C_F$ is too small (or null), the frequency response is not flat over the detector's BW and features a gain increase (similar to a peak) at frequencies close to $f_{TIA}$. The higher the peak, the higher the chances for the detector to saturate. 
In contrast, increasing the value of $C_F$ to a few picofarads can help stabilize the circuit by dampening the peak at high frequencies. However, if $C_F$ is too high, then the detector BW will decrease all together. Therefore, it is important to fine tune $C_F$ to produce a stable detector without truncating the maximum achievable BW. 

In this paper, we present three models of homodyne detectors using an OPA856 operational amplifier from Texas Instruments. This OA was specifically designed for transimpediance applications over a wide bandwidth. Compared to the usually chosen OPA847 \cite{chi2011balanced, Masalov2017} or the high-bandwidth OPA855 \cite{OPA855specs,Tasker2021,Tasker2022} (Table \ref {tab:OA_table}), the OPA856 is more robust against saturation and can be used at unity gain $G_{NG,min}$, which allows to build TIA with high resistance values. Moreover, the resilience to saturation of the OPA856, implies that it can be wired with a limited number of additional electronic components, thus limiting to electronic noise of the device. 

\subsection{Signal Demultiplexing}
The signal at the output of the TIA is directed to the third block of the detector, where the DC and AC components of the signal are separated (Fig. 
\ref{fig:basicHDscheme}). The DC signal is used to align the detector, while the AC signal -- at much lower amplitude -- contains the signal of interest. This way, it is possible to amplify the quantum signal separatly from the DC without saturating the detector. Thereby, we retrieve the low-frequency signal with a low-pass filter (LPF) and the high-frequency part with a high-pass filter (HPF). We choose the lower cutoff frequency of the HPF such that we eliminate most of the noise at low frequencies while being able to discriminate the signal coming from different pulses at the AC output when we send pulsed light on the detector.

\subsection{Noise Management}
In our detectors, we use capacitors to smooth out the output of the power supplies in order to deliver a nearly constant voltage to the electronic components. Indeed, the output signal of a power supply has a DC value, which is the desired voltage and an AC part called ripple voltage, of the order of 0.5 mV$_{rms}$ between 5 Hz and 1 MHz for a GPS-1830D power supply \cite{Mazda1989}. Capacitors help reducing the amount of ripple in the voltage signal and their value depends on the amplitude and frequency of the ripple noise of the power supply. To perform efficient filtering, we use two capacitors in parallel and we place the highest capacitance right at the output of the supply. The lowest capacitance allows to get rid of residual noise. The same scheme is used for every supply in our circuits.  It is possible to combine the filter capacitors with resistors and inductors to improve the filtering, but it should be noted that they increase the overall impedance of the circuit and contribute to increase the thermal noise of the circuit. The best solution is to work with low-input-noise power supplies. 

\setlength{\arrayrulewidth}{0.3mm} 
\renewcommand{\arraystretch}{1.2} 
\begin{table}
\begin{center}
\begin{tabular}{|c|c|c|c|}
\hline 
\rowcolor{lightgray}
 & \textbf{OPA847} & \textbf{OPA855} & \textbf{OPA856}\\
\hline
GBP & 3.9 GHz & 8 GHz & 1.1 GHz \\
\hline
 $C_{A1}$ & 3.7 pF & 0.8 pF & 1.1 pF \\
\hline
$G_{NG,min}$ & 12 V/V & 7 V/V & 1 V/V\\
\hline
$f_{max}$ & 325 MHz & 1.1 GHz & 1.1 GHz \\
\hline
\end{tabular}
\caption{\small{Comparison of the specifications of three operational amplifiers (OA). GBP: Gain-Bandwidth Product;  $C_{A1}$: Input Capacitance; $G_{NG,min}$: Minimum Noise Gain; $f_{max}$: Maximum Bandwidth}}
\label{tab:OA_table} 
\end{center} 
\end{table}

\section{Results}
We present three detectors built according to the design described in the previous section. The electronic circuit is printed on a 2-layer printed circuit board (PCB), with the top layer containing the electronic-signal tracks as well as the noise-management circuit. 

\subsection{Near-Infrared Homodyne Detector}
    \subsubsection{Lower-bandwidth homodyne detector}
\begin{figure}[h]
\centering
  \subfigure[]{\includegraphics[width=0.97\linewidth]{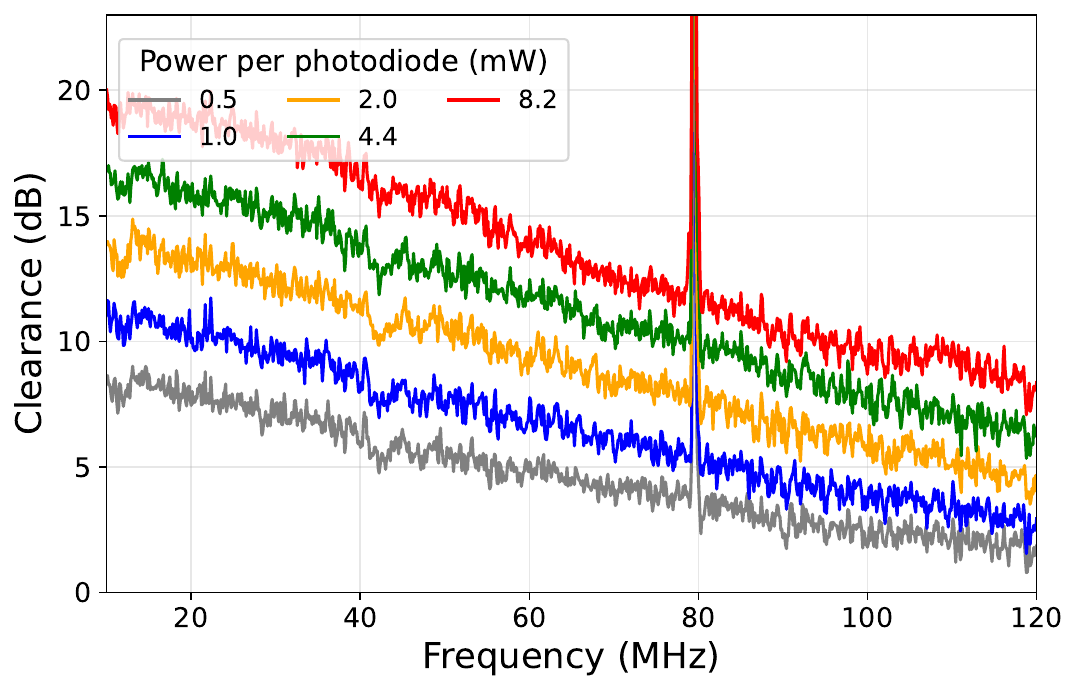}}
    \label{fig:SPOPO_clearance}
  \hfill   
    \subfigure[]{\includegraphics[width=0.97\linewidth]{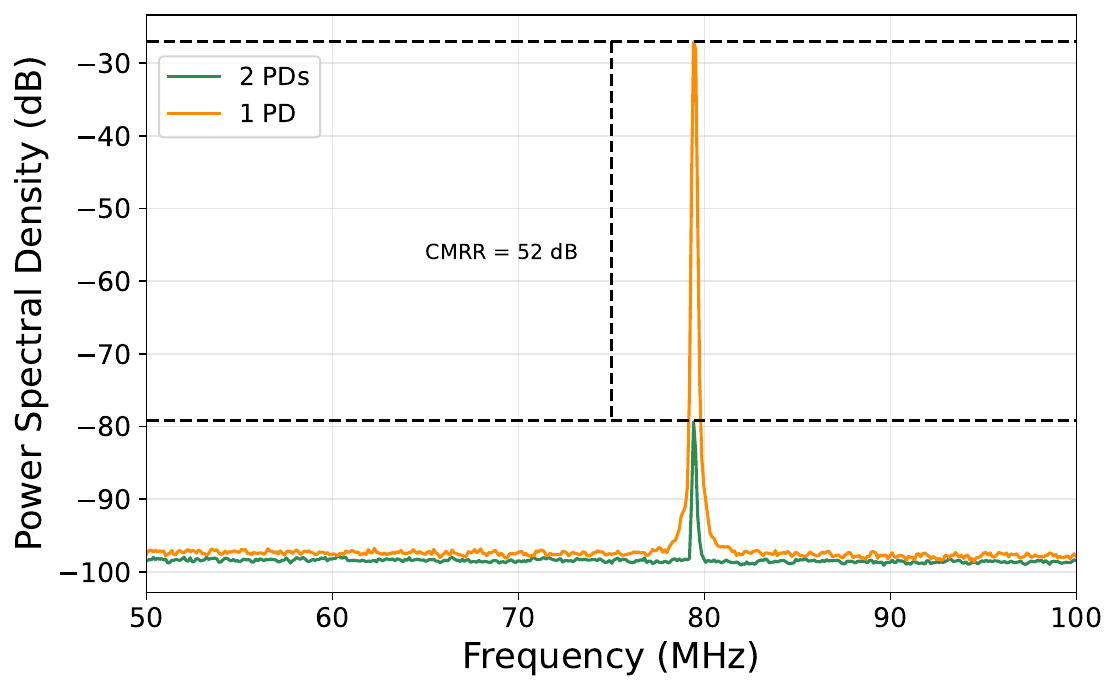}
    \label{fig:SPOPO_cmrr}}
\caption{\small{OPA 856 lower-bandwidth. (a) SNC of the NIR homodyne detector as a function of frequency. (b) Experimental measurement of the CMRR.}}
\label{fig:SPOPO_hd}
\end{figure}

This detector uses s3883 photodiodes from Hamamatsu featuring a responsivity of 0.58 A/W  and a 300 MHz bandwidth, suitable for the detection of our ultrafast 795-nm light. An OPA856 is used in a TIA configuration with a feedback resistance $R_F = 1.5 \, \text{k}\Omega$ and capacitance $C_F = 1.5 \, \text{pF}$. The signal at the output of the TIA is separated in two components, a LPF ($f_{3dB}$ = 60 kHz) isolates the signal used to align the detector and an active HPF ($f_{3dB}$ = 100 kHz) filters out the highest frequency components of the signal and further amplifies them. The AC signal is sent to an electronic spectrum analyzer (ESA) to display the power spectral density (PSD). Fig. \ref{fig:SPOPO_clearance} shows measurements of the detector's shot-noise clearance over 100 MHz and for different input powers. The detector displays a 50 MHz bandwidth and reaches a SNC of 18 dB corresponding to an effective quantum efficiency $\eta_{elec}$ of 98.4\%, without accounting for eficiency of the PDs. The detector CMRR is around 52 dB as shown in Fig. \ref{fig:SPOPO_cmrr}.

\subsubsection{Wideband NIR homodyne detector}
\begin{figure}[h]
    \subfigure[]{{\includegraphics[width=0.97\linewidth]{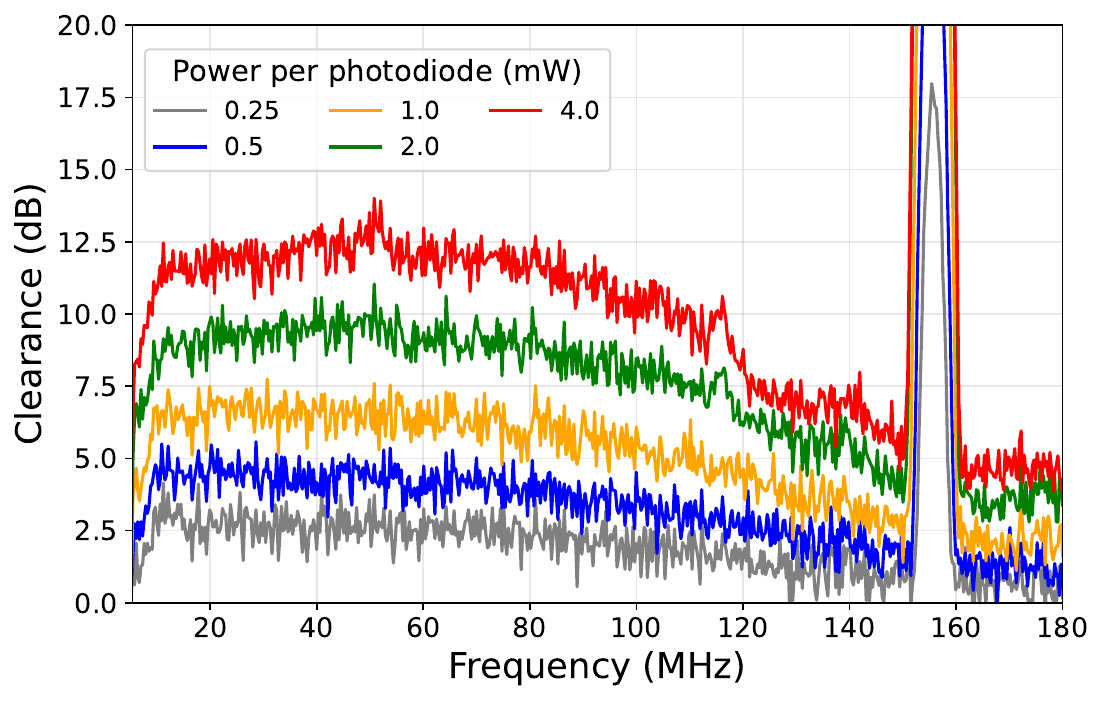} \label{fig:musiqs_hd_aa}}}
    \hfill
    \subfigure[]{\includegraphics[width=0.97\linewidth]{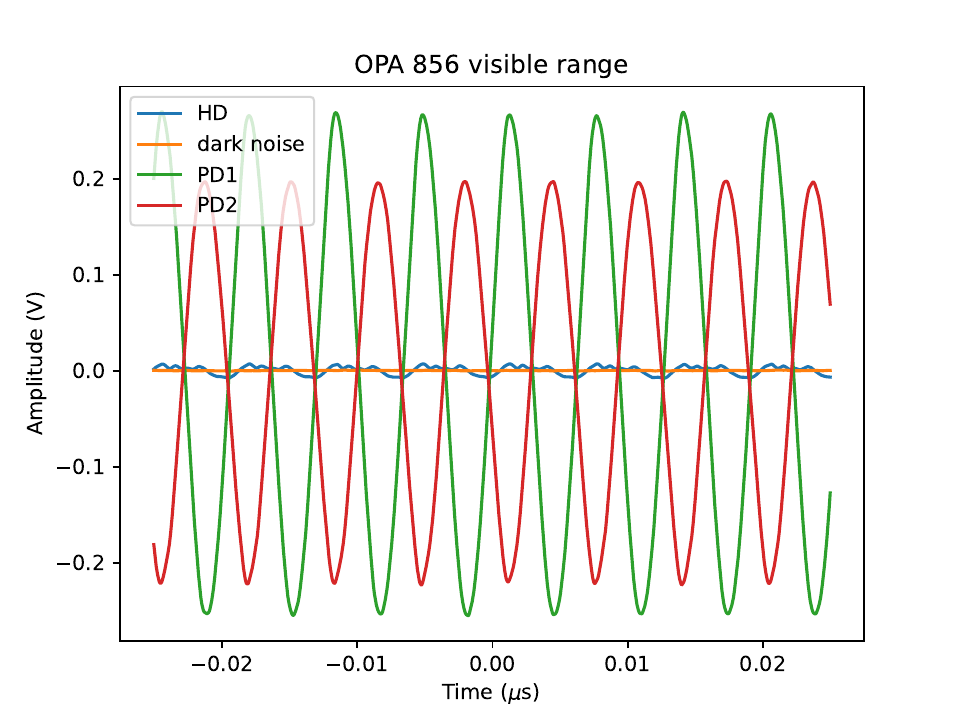} \label{fig:musiqs_hd_bb}}
\caption{\small{OPA856 Visible Range. (a) SNC of the wideband NIR homodyne detector as a function of frequency. (b) Temporal response of the detector.}}
\label{fig:musiqs_hd}
\end{figure}

The design of the second detector focuses on the bandwidth. We use the same OA and a lower feedback resistance $R_F = 620 \, \Omega$. The feedback capacitance $C_F$ of 1 pF contributes to stabilizing the response of the detector over the targeted bandwidth. As can be seen in Fig. \ref{fig:musiqs_hd_aa}, there is a trade-off between enhancing the detector gain, i.e., to increase the SNC, and the achievable bandwidth. Here, the cutoff frequency is above 100 MHz (i.e, doubled that of the previously described detector), and the highest SNC, measured at 50 MHz, is 12 dB for 4 mM of input power, i.e., a 6-dB reduction compared to the previous detector. This corresponds to a 93.6\% detector efficiency at 50 MHz. 
In \cite{kouadou2023}, we showed the measurement of multiple squeezed spectral modes using this detector design. Around 150 MHz, the SNC is 7 dB (i.e., $\eta_{elec} = 80.1 \%$).



To assess the HD performance at higher frequencies, we measure the homodyne signal, i.e., the voltage at the output of the HD, with a fast oscilloscope (WaveRunner 8404M). The results are shown in Fig. \ref{fig:musiqs_hd_bb}. The green and red curves correspond to the measured output voltage when only PD1 (PD2) is illuminated. The sinusoidal shape of the temporal signal corresponds, in the spectral domain, to the CM peak at the laser repetition rate (visible in Fig. \ref{fig:musiqs_hd_aa}). The orange and blue curves are, respectively, the dark noise and the output voltage when both PDs are illuminated at the same time (homodyne signal). The amplitude difference between the blue and the green (red) curve illustrates the CMRR, i.e., the ability of the HD to subtract the photocurrents $i_1$ and $i_2$. The CMRR of this detector is 62 dB. 

\subsection{Telecom-range homodyne detection}
\subsubsection{Wide C-band detector} \label{Wide C-band detector}

\begin{figure}[h]
\centering
\includegraphics[width=8.5cm]{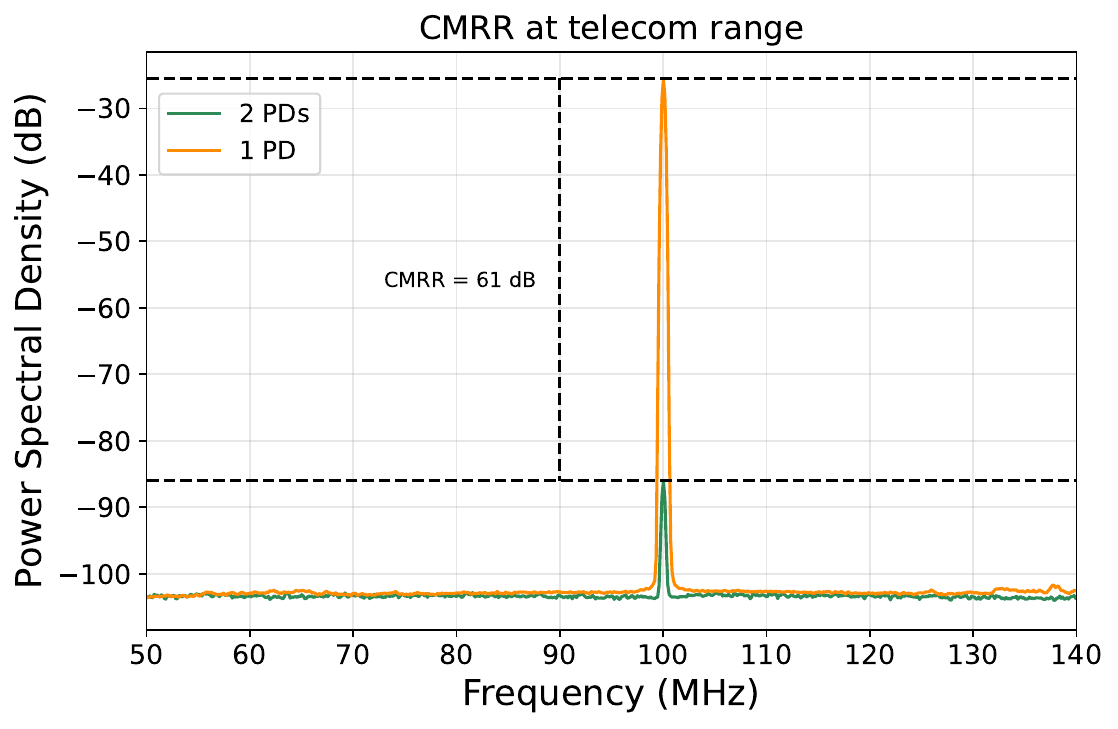}
\caption{\small{Measurement of the CMRR of the telecom detector at 65~$\mu $W of input optical power. The high-amplitude signal at ~100 MHz is the repetition rate of the laser. The electronic signal from one photodiode (orange) is compared to the same signal when both PD are illuminated (green).}
\label{fig:muteq_cmrr_144}}
\end{figure}

\begin{figure}[h]
\centering
  \subfigure[]{\includegraphics[width=0.97\linewidth]{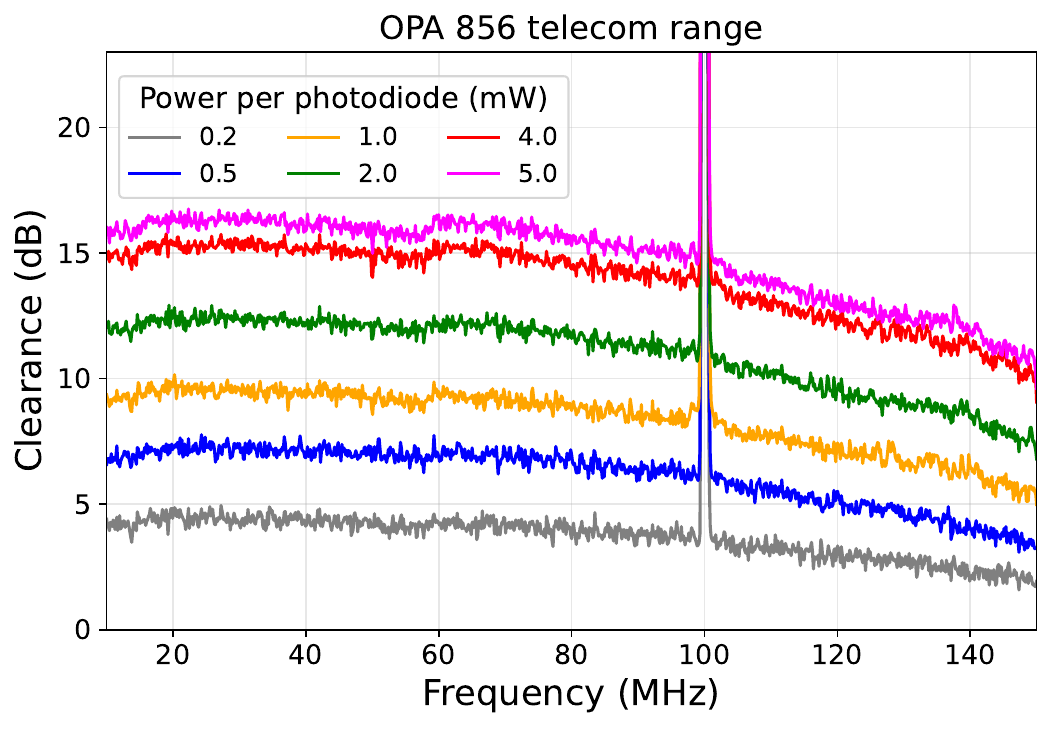}
    \label{fig:muteq_snc_aa}}
    \hfill
  \subfigure[]{\includegraphics[width=0.97\linewidth]{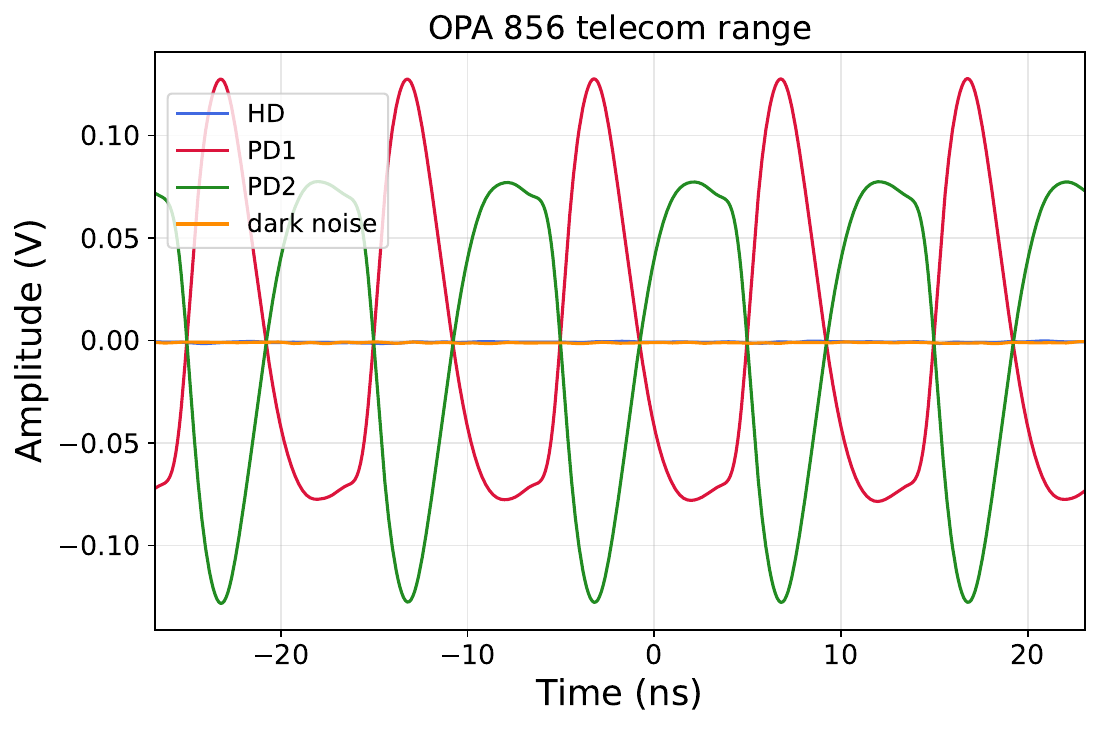}
    \label{fig:muteq_snc_bb}}
\caption{OPA856 Telecom Range. (a) SNC measurement of the telecom HD. (b) Temporal response of the detector.}
\label{fig:muteq_snc}
\end{figure}
 
Our telecom source is a mode-locked C-Fiber Femtosecond Erbium laser centered at 1560 nm, and emiting 64 fs pulses at a 100 MHz repetition rate. For the homodyne detection of telecom-range squeezed light, we use the same detector design
and replace the Si PD with InGaAs ones. 

We use Thorlabs FGA015 InGaAs photodiodes featuring a 0.95 A/W responsivity at 1550 nm and 3-GHz BW. It should be noted that over their detection range, the responsivity of InGaAs photodiodes increases with the wavelength (eq. \ref{eq:QE_responsivity}) and reaches a maximum around 1560 nm \cite{FGA015specs}. For this reason, these PDs have a lower saturation threshold than the s3883 PDs used in Section 3.A, therefore reducing the saturation threshold of the overall detector. 

To try to overcome the saturation issues, we set the feedback parameters of the TIA to $R_F = 1.2 \,\text{k}\Omega$ and no capacitor; this extends the BW of the HD to frequencies beyond the laser common mode, while filtering out the signals outside the range of squeezed-light production. Additionally, in order to optimize the subtraction of photocurrents before the TIA, i.e. the detector CMRR, we built a separate electrical circuit to characterize the PDs. We measure their intrinsic capacitance, and select pairs whose relative capacitances can be equalized with the voltage biasing. In this configuration, we measured a CMRR of 61 dB (Fig. \ref{fig:muteq_cmrr_144}). 

In Fig. \ref{fig:muteq_snc}.(a), we present the measurements of the SNC over the detector's bandwidth and for multiple input powers. We can see that over 100 MHz, the detector clearance increases with the power. In particular, at $\text{P}_{in} = 4$~mW  the SNC reaches 15 dB over an 80-MHz range before saturation. This corresponds to an overall effective quantum efficiency of $\eta_{elec}=$ 96.84\%. Figure \ref{fig:muteq_snc}.(b) shows the temporal response of each photodiode. The periodicity of the signal is 10~$ns$, corresponding to the repetition rate of the laser. Moreover, it shows that the designed HD is capable of resolving pulse by pulse signals.

\begin{figure}[h]
  \subfigure[]{\includegraphics[width=0.97\linewidth]{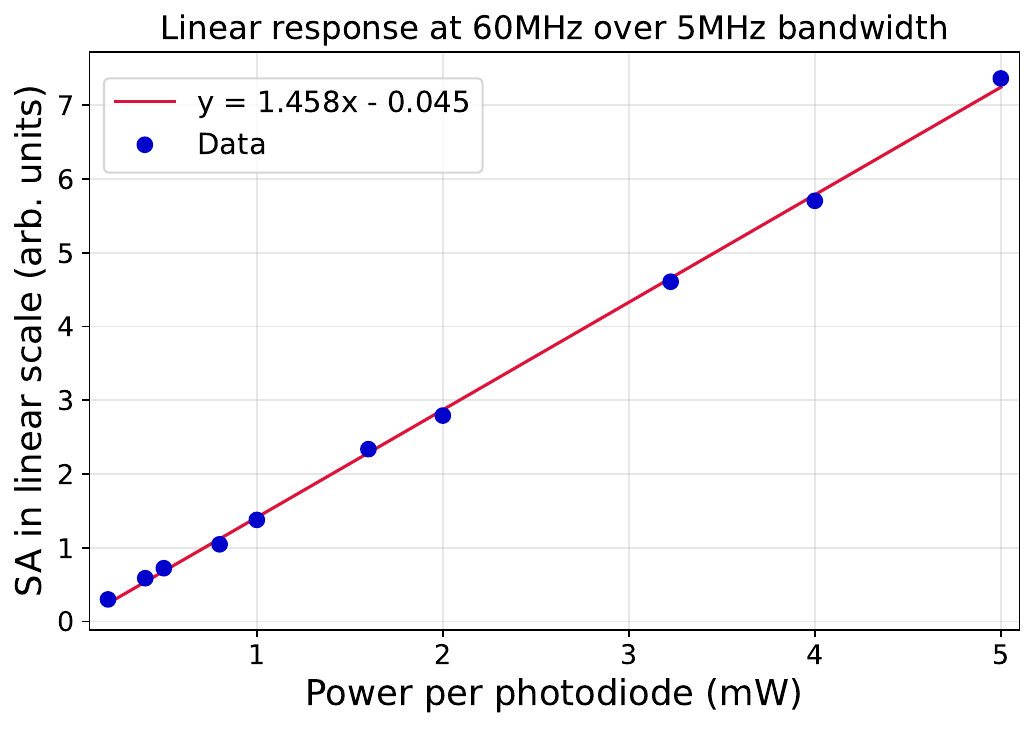}}
   \hfill
  \subfigure[]{\includegraphics[width=0.97\linewidth]{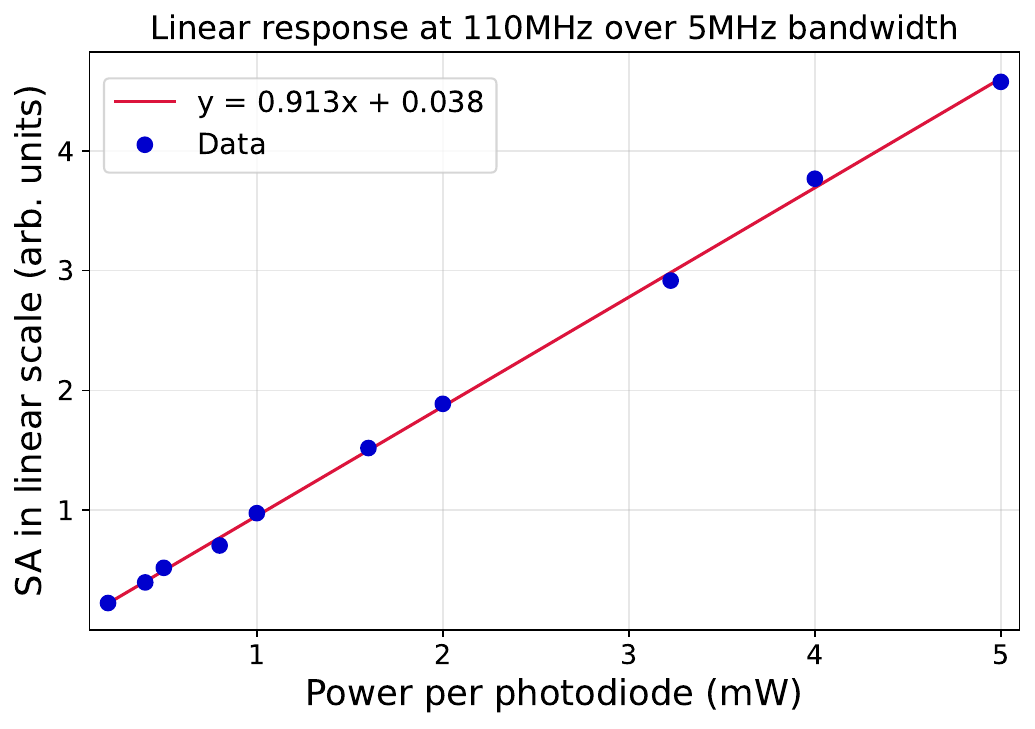}}
  \caption{Response of the noise signal as a function of incoming power. The noise level corresponds to the linearized spectrum analyzer signal. Top-figure: analysis frequency centered at 60 MHz. Low-figure: analysis frequency equal to 110 MHz. The data was taken using 5 MHz bandwidth. }
\label{fig:det_response}
\end{figure}

To evaluate the saturation threshold of the telecom HD, we measured the output PSD as a function of the LO power (Fig. \ref{fig:det_response}). Each curve shows measurements of the detector SNC at an analysis frequency centered around 60 MHz and 110 MHz, over a span of 5 MHz and between 36 $\mu$W and 9 mW of optical power. At different frequencies along the detector's BW, we see that its response remains linear with a slope around 3 dBm/mW. On the contrary, if the detector approaches saturation, the slope of this function will decrease until reaching zero, i.e., the measured signal will stop evolving with the input optical power. This measurement allows us to assess the bandwidth over which the HD is operational; for pulse-by-pulse squeezing measurements, the HD response must remain linear around the laser repetition rate, i.e., 100 MHz, which is the case. Moreover, the highest (non-saturating) input power sets the highest measurement clearance, i.e., the power of largest signal-to-noise ratio (SNR).

In the next section, we present measurement of multimode squeezed states at telecom wavelengths using this detector.

\subsection{Pulse by pulse multi-spectral squeezing measurements}
To test the on-field performance of the detector, we used it for the detection of multiple squeezed optical spectra associated with the pulse-by-pulse generation of squeezing at telecom wavelengths.
This regime presents particular challenges because photodiodes operating at these wavelengths exhibit reduced efficiency ($\eta_{PD}$), and the very broad optical spectra are not always perfectly mode-matched to the shaped local oscillator in homodyne detection, thereby reducing the mode-matching efficiency ($\eta_{mod}$). Consequently, to accurately capture the quantum features of the signal, the electronic efficiency ($\eta_{elec}$) must be preserved, avoiding significant degradation due to the gain–bandwidth trade-off when fast detectors are used to resolve the pulse-by-pulse structure.

\begin{figure}[h]
\includegraphics[width=\linewidth]{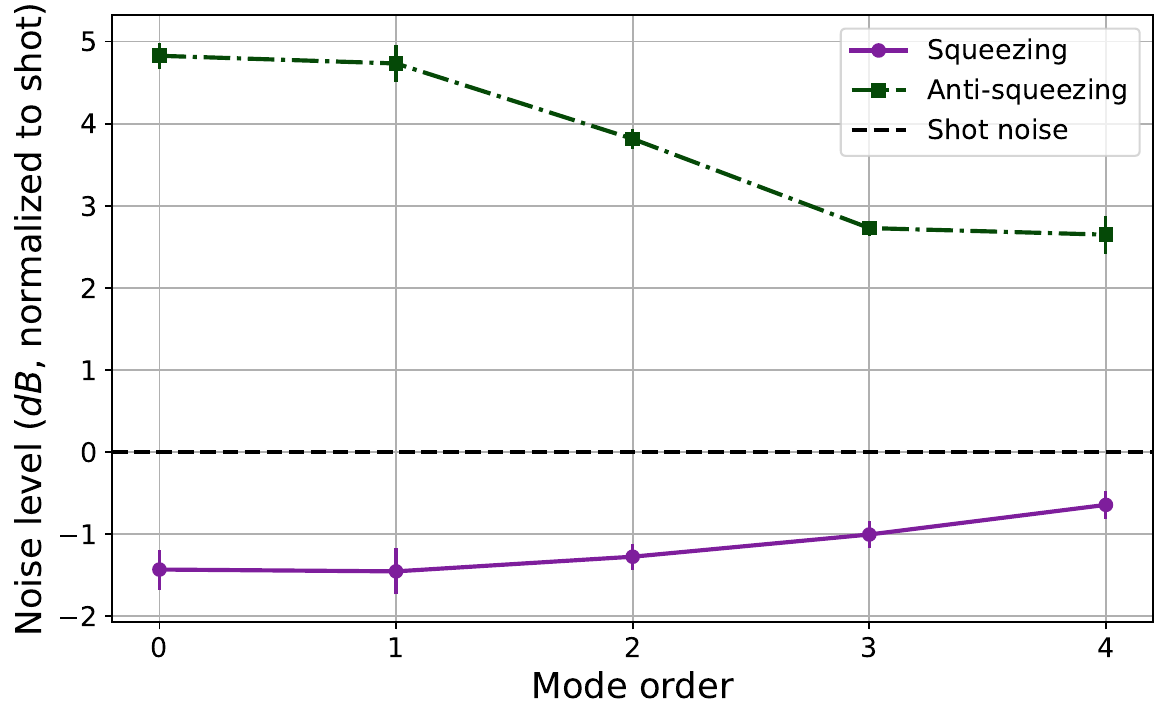}   
\caption{Pulse-by-pulse squeezing measurements. The data was acquired using a fast oscilloscope that resolved 10 G samples per second. The quadrature's noise level was obtained in postprocessing, by integrating on a 10 ns time window and calculating the variance of the quadratures.
\label{fig:pulse_squeezing_meas}}
\end{figure}

We measure the quadratures of multimode squeezed states generated via degenerate type-0 spontaneous parametric down conversion (SPDC) in a periodically poled potassium titanyl phosphate (ppKTP) waveguide. The nonlinear process is pumped with the mode-locked laser described in \ref{Wide C-band detector}, after undergoing second harmonic generation (SHG) in a periodically poled lithium niobate (ppLN) crystal. This results in a train of multimode squeezed pulses centered at 1560 nm. A more detailed description of the setup can be found in \cite{RomanFainsin2024}. The down converted light exhibits a rich modal structure originating from the nonlinearity of the waveguide and the spectral width of the pump, which can be decomposed into an eigenbasis of squeezed spectral modes, called \textit{supermodes} \cite{kouadou2023}. Each spectral supermode is characterized by a squeezing eigenvalue and an associated eigenfunction \cite{Patera2010}.  Numerical calculations show that such eigenfunctions are closely related to the Hermite-Gauss basis whose elements are indicated with $\text{HG}_{n}$. 
The squeezed light is sent to a homodyne detection setup similar to the one in Fig. \ref{fig:basicHDscheme}. The phase of the LO is scanned with a piezoelectric actuator to access different quadratures, and its spectrum is shaped with an ultrafast pulse shaper \cite{Weiner2011,Monmayrant2010} to select the mode to be measured. A detailed description of the pulse shaping technique is available in \cite{RomanFainsin2024}.  The current optical setup improved the wavelength range that can be shaped going from 60 nm of \cite{RomanFainsin2024}  to more than 90 nm.
The Hermite-Gauss  basis used here correspond to the one where the mode $\text{HG}_{0}$ has a Full Width Half Maximum of 25 nm, such width has been chosen by maximizing the amount of squeezing of this first mode.

To access the temporal structure of the squeezed state, we perform pulse-resolved homodyne detection. The homodyne signal is sent to a fast oscilloscope (WaveRunner 8108HD) with sufficient bandwidth to resolve individual femtosecond-scale pulses. 
A quadrature value $q_{\theta}(t)$ is obtained for each pulse by integrating the signal over a carefully chosen temporal window, ensuring separation between consecutive pulses. Following the protocol in \cite{kouadou2023},
we collect traces over 1 ms intervals, corresponding to $10^5$ pulse slots. In each slot, 100 oscilloscope data points are integrated to obtain a quadrature value for the pulse. We then extract the variance of the resulting set of quadratures to determine the squeezing values.
 In Fig.~\ref{fig:pulse_squeezing_meas} we show the results of the pulse-resolving characterization for 5 different spectral modes, from $\text{HG}_0$ to $\text{HG}_4$. The dashed line corresponds to the shot noise level; below it the quadratures are squeezed while above it they are anti-squeezed. Notice that it is expected that squeezing and anti-squeezing decrease as the mode-order increases, as has been shown in \cite{RomanRodriguez2024}. While a full analysis of the multimode spectral content of the quantum resource is beyond the scope of this article, the pulse-by-pulse detection of squeezing in several modes of the Hermite–Gauss series demonstrates that the homodyne detector can be exploited in protocols requiring multiplexing both in the time-bin (pulse) and optical spectral domains \cite{kouadou2023}.

\section{Discussion}
We presented the development of three homodyne detectors using the same circuit design based on the OPA856 operational amplifier. After introducing the principles of homodyne detection and the building blocks of a balanced homodyne detector, we demonstrated the implementation of our design to different wavelengths, i.e., NIR and telecom, with comparable performance, and we optimized the gain-bandwidth product of the circuits to produce detection bandwidths between 60 and 110 MHz and shot-noise clearances between 7 and 18 dB. We demonstrated the ability of our design to characterize ultrafast squeezed-light sources at NIR \cite{kouadou2023} and telecom wavelengths. Our detectors can withstand kilowatts of peak power, and exhibit over 50 dB CMRR. This is especially beneficial in the pulse-by-pulse characterization of current sources of spectrally multimode squeezed states.

We emphasize the versatility of our design in the characterization of quantum states in  a variety of settings, namely, continuous wave (CW) and mode-locked, high-frequency, and wide-spectrum signals. 

\section{Acknowledgments}
This   work   was   supported   by   the   European   Research Council under the Consolidator Grant COQCOoN (Grant No.  820079) and by Agence Nationale de la Recherche (OQuLus (ANR-22-PETQ-0013)). P. N. was supported by QuantEdu-France (ANR-22-CMAS-0001) in the framework of France 2030.

%
\end{document}